\documentclass[prl,twocolumn,floats,aps,epsfig,nofootinbib,amssymb]{revtex4}
\usepackage{subfigure}
\usepackage[dvips]{graphicx}
\usepackage{amsmath}
\usepackage{bm}
\usepackage{times}
\usepackage{epsfig}

\newcommand{\be}{\begin{equation}}
\newcommand{\ee}{\end{equation}}
\begin{document}
\title{\hspace{-0.75cm} Left-right models with light neutrino mass prediction 
and dominant neutrinoless double beta decay rate}
\author{\bf M. K. Parida}
\email{paridamk@soauniversity.ac.in}
\affiliation{Center of Excellence in Theoretical and Mathematical Sciences, \\
Siksha 'O' Anusandhan University, Bhubaneswar-751030, India}
\author{\bf Sudhanwa Patra}
\email{sudha.astro@gmail.com}
\affiliation{Center of Excellence in Theoretical and Mathematical Sciences, \\
Siksha 'O' Anusandhan University, Bhubaneswar-751030, India}
\begin{abstract}
In TeV scale left-right symmetric models, new dominant predictions to neutrinoless double 
beta decay and light neutrino masses are in mutual contradiction because of large contribution 
to the latter through popular seesaw mechanisms. We show that in a class of left-right models 
with high-scale parity restoration, these results coexist without any contravention with neutrino 
oscillation data and the relevant formula for light neutrino masses is obtained via gauged inverse 
seesaw mechanism. The most dominant contribution to the double beta decay is shown to be via 
$W^-_L- W^-_R$ mediation involving both light and heavy neutrino exchanges, and the model predictions 
are found to discriminate whether the Dirac neutrino mass is of quark-lepton symmetric origin or 
without it. We also discuss associated lepton flavor violating decays. 
\end{abstract}
\maketitle
{\hspace{-0.35cm}\bf I. INTRODUCTION:~}
Evidences of tiny neutrino masses uncovered by the solar, atmospheric, and reactor neutrino 
oscillation experiments while calling for physics beyond the Standard Model (SM) might be 
strongly hinting at the fundamental nature of the particle i.e. whether Dirac \cite{Dirac} 
or Majorana \cite{Maj}. In fact popular theories based upon seesaw mechanisms like type-I 
seesaw \cite{seesaw1}, type-II \cite{type-IImoh, seesaw2}, type-III \cite{type-III, goran-typeIII}, 
inverse seesaw \cite{invo, inv, invdet,  inverseso10}, and others \cite{tevseesaw, inverseothers, 
Kang-Kim, Parida} come out with natural predictions of light Majorana neutrino masses.  
With lepton number violating mass insertion term by two units, confirmation of any events 
at the experimental search programmes for the neutrinoless double beta decay ($0\nu 2\beta$) 
would not only indicate the Majorana nature of the particles, but also it would 
strongly support the underlying seesaw mechanism for their mass generation. There have been 
attempts \cite{epj, igex,Gerda, Cuore, exo} on the experimental side to observe 
such a rare process, even with a present claim \cite{klapdor} while others are trying to improve 
the life time for this $0\nu 2 \beta$ process \cite{cuoricino,nemo,supernemo}.
So far, the Heidelberg-Moscow experiment using $^{76}{Ge}$ \cite{klapdor} has given the best 
limit on the half life, $T_{1/2} < 3 \times 10^{25}$ Yrs. which gives an upper bound on effective 
neutrino mass, $m_{\mathrm{eff}} \leq 0.21 - 0.53~ {\rm eV}$. 
There are several interesting discussions and models using different seesaw mechanisms \cite{hirsch, 
allanach, vogel, choi, tello-senj, Ibarra, Mitra:2011qr, Ibarra:2011xn, spatra-jhep, blennow} 
exploring possible non-standard contributions to $0\nu 2\beta$ transition.

Another important mysterious phenomenon of SM, namely, the origin of parity 
violation as monopoly of weak interactions, has been suggested to be having 
its underlying origin in the left-right (LR) symmetric interactions \cite{yang} 
which could be through the existence of mirror particles \cite{mirror} of the 
SM or via left-right symmetric gauge theories \cite{pati, moh-senj}. A very 
attractive aspect of LR gauge theory is its potential to explain the origin 
of parity ($P$) and $CP$ violations in weak interactions and small neutrino 
masses. If left-right gauge theory has to make any significant impact on weak 
interactions phenomenology, the associated $W^\pm_{R}$ and $Z_{R}$ boson masses 
have to be low. While current searches at the Large Hadron Collider restricts 
the lower bound on the scale of the RH gauge boson masses ($M_R$) to be $\mathcal{O} 
(1)$ TeV, $K_L-K_S$ mass difference gives $M_R > 2.5$ TeV \cite{WR-limit}. Such a low scale 
$W_R$ boson associated with right-handed charged currents can give additional 
contributions to $0\nu 2 \beta$ and can be also accessible to LHC and future 
accelerator searches. As a result of this, there can be various non-standard 
contributions to $0\nu 2\beta$ in LR gauge theories mediated by: 
({\bf 1}.) two $W_L$ gauge bosons (associated with left-handed currents), 
({\bf 2}.) two $W_R$ gauge bosons (associated with right-handed currents), 
({\bf 3}.) one $W_L$ and one $W_R$ gauge boson at each vertex (mixed diagram) 
accompanied by both light and heavy neutrinos \cite{tello-senj,spatra-jhep}. 
In addition, there could be other contributions to $0\nu 2\beta$ in LR model due to 
doubly charged Higgs scalar exchanges where Majorana neutrino mass insertion 
has no role to play \cite{note}. It is important to note here that the contributions 
to $0\nu 2\beta$ from the mixed diagram has been either ignored or considered to be 
sub-dominant \cite{Miha}, although this has been taken into account in the inverse process $e^- e^- 
\rightarrow W^-_L W^-_R$ in Ref. \cite{rode-ilc} for linear collider searches. 

The natural TeV mass scale for RH Majorana neutrinos in conventional low scale LR 
gauge models emphasizing upon light neutrino mass generation mechanisms however 
predicts very large contribution to the light neutrino masses through canonical 
or type-II seesaw mechanisms \cite{type-IImoh,moh-senj,pati}. Thus, it turns out that 
new dominant contributions to observable neutrinoless double beta decay ($0\nu 2\beta$) 
can not coexist with the experimentally determined tiny neutrino masses \cite{DAYA-BAY}. Alternatively, 
interesting proposals have been advanced where type-II seesaw dominance \cite{tello-senj,spatra-jhep} 
has been invoked by suppressing Dirac neutrino mass matrix in which case LR gauge theories 
may have only sub-dominant roles to play in representing charged fermion masses. The purpose 
of this letter is two fold: while showing that a crossed diagram with simultaneous 
$W^-_L$ and $W^-_R$ exchanges predicts the most dominant contribution to the ($0\nu 2\beta$), 
we provide a class of TeV scale left-right gauge theories where this is implemented 
without any suppression of naturally permitted Dirac neutrino masses and 
without any contravention with the neutrino oscillation data. The neutrino mass generation 
mechanism in these models turns out to be through gauged inverse seesaw.

{\hspace{-0.37cm}\bf II. THE MODEL:~}
In conventional LR gauge theories, the type-I \cite{seesaw1} and type-II seesaw \cite{type-IImoh} 
contributions to light neutrino masses are 
\begin{equation*}
m^{\rm I}_\nu \sim - M_D\, \frac{1}{M_N}\, M^T_D\hspace{0.5cm}, \hspace{0.5cm} m^{\rm II}_\nu = f\, v_L\, 
\end{equation*}
where the Dirac neutrino mass matrix $M_D$ is 
similar to the charge lepton mass matrix, or the up-quark mass matrix if the model has 
its origin from Pati-Salam symmetry. The induced triplet vacuum expectation value is 
$v_L = \lambda_{\rm eff}\, v^2_{\rm wk}/M_{\Delta_L}$. Then the natural seesaw scales 
consistent with neutrino oscillation data are $M_N \geq (10^{11}-10^{14})$ GeV and the 
TeV scale LR gauge models relevant for $0\nu 2\beta$ are ruled out. We now construct a 
class of LR gauge models where $W^\pm_R$ and $M_N$ are allowed near the TeV scale which 
contribute predominantly to $0\nu 2 \beta$, yet the model does not upset small neutrino 
mass predictions consistent with the neutrino oscillation data. In our model although 
the parity restoration scale is large, yet the asymmetric left-right (LR) gauge theory 
$SU(2)_{L} \times SU(2)_{R} \times U(1)_{B-L} \times SU(3)_{C}\, ~[\equiv \mathcal{G}_{2213}]$ 
($g_{2L} \neq g_{2R}$) survives down to the TeV scale subsequent to the D-parity breaking 
\cite{chang}. 
To implement the idea we use the set of Higgs scalars with their gauge quantum numbers 
under $\mathcal{G}_{2213}$ $\sigma (1, 1, 0, 1) \oplus \Delta_{L} (3, 1, -2, 1) \oplus 
\Delta_{R} (1, 3, -2, 1), \chi_{L} (2, 1, -1, 1) \oplus \chi_{R} (1, 2, -1, 1) \oplus 
\Phi (2, 2, 0, 1)$ where $\sigma$ is D-parity odd. It is well known that by assigning 
large parity breaking vacuum expectation value (vev); $\langle \sigma \rangle \sim M_P$, 
the model gives all the left-handed (LH) Higgs scalars to have heavy masses i.e. 
$\mathcal{O}(M_P)$ while those of the right-handed (RH) scalars can have much lighter 
masses near the TeV scale with $M^2_{\Delta_R} \simeq \left(\mu^2_{\Delta_R} -\lambda 
\langle \sigma\rangle M \right)$ and $M^2_{\chi_R} \simeq \left(\mu^2_{\chi_R} -\lambda^\prime 
\langle \sigma\rangle M \right)$ where $M_{\Delta_L} \sim M_{\chi_L} \sim M \sim \mathcal{O} 
(M_P)$. In fact $M_{\Delta_R}$ and $M_{\chi_R}$ can have any value below $M_{P}$ depending 
upon the degree of fine tuning in $\lambda$ and $\lambda^\prime$. The asymmetry in the 
Higgs sector at the energy scales below $\mu \sim M_P$ causes asymmetry in the gauge 
couplings, $g_{2L} \neq g_{2R}$ for the surviving left-right gauge group. Alternatively, 
the asymmetric LR model may emerge from high scale Pati-Salam symmetry $SU(2)_L \times SU(2)_R 
\times SU(4)_C \times D$ ($g_{2L}=g_{2R}$) with similar choice on the Higgs scalars. 
In particular, we examine the TeV scale phenomenology for neutrino masses and $0\nu 2\beta$ 
with the following two possible cases of symmetry breaking:

{\bf A:}
\begin{eqnarray}
& &SU(2)_L \times SU(2)_{R} \times U(1)_{(B-L)} \times SU(3)_C \times D ~~\nonumber \\ 
& &\hspace{3.5cm} [\equiv \mathcal{G}_{2213D}]~~ (g_{2L} = g_{2R})  \nonumber \\  
& &\hspace{0.2cm} \stackrel{M_P}{\longrightarrow} 
        SU(2)_L  \times SU(2)_R \times U(1)_{(B-L)} \times SU(3)_{C} \nonumber \\
& &\hspace{3.5cm} [\equiv \mathcal{G}_{2213}]~~ (g_{2L} \neq g_{2R})  \nonumber \\ 
& &\hspace{0.2cm} \stackrel{M_R}{\longrightarrow} 
         ~~~G_{SM}
\end{eqnarray}

{\bf B:}
\begin{eqnarray}
& &SU(2)_L \times SU(2)_{R} \times SU(4)_C \times D \nonumber \\ 
& &\hspace{3.5cm} [\equiv \mathcal{G}_{224D}]~~ (g_{2L} = g_{2R})  \nonumber \\ 
& &\hspace{0.2cm} \stackrel{M_P}{\longrightarrow} 
SU(2)_L \times SU(2)_{R} \times U(1)_{(B-L)} \times SU(3)_C  \nonumber \\
& &\hspace{3.5cm} [\equiv \mathcal{G}_{2213}]~~ (g_{2L} \neq g_{2R})  \nonumber \\ 
& &\hspace{0.2cm} \stackrel{M_R}{\longrightarrow} 
         ~~~G_{SM}
\end{eqnarray}
One important difference between the two scenarios is that in model-A, the Dirac neutrino 
mass matrix is similar to the charged lepton mass matrix while in model-B, it is similar 
to up-quark mass matrix.

In addition to the standard $16$-fermions of each generation, we require one additional 
fermion singlet for each generation ($S_i$, i=1, 2, 3) which is essential for the 
implementation of inverse seesaw mechanism \cite{inv} or, the so called extended 
seesaw mechanism \cite{Kang-Kim,ellis, Parida}. The renormalizable Yukawa Lagrangian 
near the TeV scale with asymmetric LR gauge theory then turns out to be
\begin{eqnarray}
\mathcal{L}_{\rm Yuk} &= & Y^{\ell} \overline{\psi}_L\, \psi_R\, \Phi 
                       + f\, \psi^c_R\, \psi_R \Delta_R 
                       + F\, \overline{\psi}_R\, S\, \chi_R \nonumber \\
                       &+& S^T \mu_S S +\text{h.c.} 
\end{eqnarray}
where $\mu_S$ is the singlet fermion mass matrix. We break the LR gauge theory spontaneously 
to SM by the vev $\langle \Delta^0_R \rangle =v_R$ ($\simeq M_R$) while the vev $\langle \chi^0_R \rangle 
= v_\chi$ ($\leq M_R$) is used to generate the $N-S$ mixing. The SM breaks to the low energy 
symmetry by the VEV of the SM Higgs doublet in $\Phi$. With this structure of the Yukawa 
Lagrangian, the full ($9 \times 9$) neutrino mass matrix in the ($\nu_L$, $N_R$, $S_L$) 
basis is given by 
\begin{equation}
\mathcal{M}= \left( \begin{array}{ccc}
                0        & M_D   & 0   \\
              M^T_D &    M_N         & M \\
              0 & M^T & \mu_S
        \end{array} \right) 
\label{eqn:numatrix}       
\end{equation}
where $M=F\, v_\chi$, $M_D = Y^{\ell}\, \langle \Phi \rangle$, and $M_N=f\, v_R$. 
Here $M_D$ and $M$ are $3\times3$ complex matrices in flavor space and $\mu_S$ is 
the $3\times 3$ complex symmetric matrix. 

For implementation of the light neutrino mass generation mechanism the desired hierarchy 
$M_N \gg M \gg M_D \gg \mu_S$ with a fine tuned small lepton number violating parameter $\mu_S$ 
can be easily satisfied in the model after spontaneous symmetry breaking. Since the right-handed 
neutrinos are assumed to be larger than other mass scales, they eventually decouple at low scales 
\cite{Kang-Kim, ellis, Parida}. It is important to note that this extended seesaw scenario is very 
different from the inverse seesaw scenario \cite{invo, inv,inverseso10} due to the 
simultaneous presence of both the heavy and small lepton number violating scales $M_N$ 
and $\mu_S$. Complete block diagonalization of eq. (\ref{eqn:numatrix}) gives 
the usual inverse seesaw formula for light neutrino masses
\begin{equation}
m_\nu = \left(\frac{M_D}{M}\right) \mu_S \left(\frac{M_D}{M}\right)^T\, ,
\label{eqn:inverse-massform}
\end{equation}
as well as the heavy neutrino mass matrices: $m_N \simeq M_N$ and $m_S \simeq M \frac{1}{M_N} 
M^T$. It is important to note that with $M_N \gg M \gg M_D, ~\mu_S$, the type-I seesaw contribution 
to the light neutrino mass matrix, i.e. $-M_D \frac{1}{M_N} M^T_D$ cancels out after complete 
block diagonalization. Also these block diagonal mass matrices $m_\nu$, $m_S$ and $m_N$ can 
further be diagonalized to give physical masses for all neutral leptons by respective unitary 
mixing matrices: $U_\nu$, $U_{S}$ and $U_{N}$ where
\begin{eqnarray}
U^\dagger_\nu\, m_{\nu}\, U^*_{\nu}  &=& \hat{m}_\nu = 
         \text{diag}\left[m_{\nu_1}, m_{\nu_2}, m_{\nu_3}\right]\, , \nonumber \\ 
U^\dagger_S\, m_{S}\, U^*_{S}  &=& \hat{m}_S = 
         \text{diag}\left[m_{S_1}, m_{S_2}, m_{S_3}\right]\, , \nonumber \\
U^\dagger_N\, m_{N}\, U^*_{N}  &=& \hat{m}_N = 
         \text{diag}\left[m_{N_1}, m_{N_2}, m_{N_3}\right]\,.
\label{eq:nudmass}
\end{eqnarray}

The relevant charged current interactions of leptons for this TeV scale LR gauge theory 
in the flavor basis is given by
{\small 
\begin{eqnarray*}
\mathcal{L}_{\rm CC} &=& \frac{g}{\sqrt{2}}\, \sum_{\alpha=e, \mu, \tau}
\bigg[ \overline{\ell}_{\alpha \,L}\, \gamma_\mu {\nu}_{\alpha \,L}\, W^{\mu}_L 
      + \overline{\ell}_{\alpha \,R}\, \gamma_\mu {N}_{\alpha \,R}\, W^{\mu}_R \bigg] 
      + \text{h.c.} 
\end{eqnarray*}
}
where, in terms of mass eigenstates ($\nu_{m_i}$, $S_{m_j}$, $N_{m_k}$) \cite{Mitra:2011qr},
\begin{eqnarray}
& &\nu_{\alpha\,L} \sim \mathcal{N}_{\alpha\, i}\, \nu_{m_i} + \mathcal{U}^{\nu\, S}_{\alpha\, j}\, S_{m_j} +
                      \mathcal{U}^{\nu\, N}_{\alpha\, k}\, N_{m_k}\, , \nonumber \\
& &N_{\alpha\,R} \sim \mathcal{V}^{N\, \nu}_{\alpha\, i}\, \nu_{m_i} + \mathcal{V}^{N\, S}_{\alpha\, j}\, S_{m_j} +
                      \mathcal{R}_{\alpha\, k}\, N_{m_k}\, , \nonumber \\
& &    \mathcal{N}_{\alpha\, i}= \{\left(1- \frac{1}{2} X\, X^{\dagger}\right)\, 
                                    U_\nu \}_{\alpha\, i}\, , \nonumber \\
& &     
        \mathcal{R}_{\alpha\, k}= \{\left(1 - \frac{1}{2}\, X''^\dagger\, X''\right)\, 
        U_N\}_{\alpha\, k} \, ,
\nonumber \\  & &
        \mathcal{U}^{\nu\, S}_{\alpha\, j} = \{-X\, U_S\}_{\alpha\, j} \, , ~~
        \mathcal{U}^{\nu\, N}_{\alpha\, k} = \{-X'\, U_N \}_{\alpha\, k}\, ,
\nonumber \\  & &
        \mathcal{V}^{N\, \nu}_{\alpha\, i} = \{X^\dagger\,\frac{\mu_S}{M}\, 
        U_\nu \}_{\alpha\, i}\, , ~~
        \mathcal{V}^{N\, S}_{\alpha\, j} = \{ - X''^\dagger\, U_S\}_{\alpha\, j}\, .
\label{eqn:nustate-mass}
\end{eqnarray}
The non-unitarity matrices in our model are $X=\frac{M_D}{M}$, $X^\prime=\frac{M_D}{M_N}$, 
and $X''=\frac{M}{M_N}$ due to $\nu-S$, $\nu-N$, and $S-N$ mixings, respectively.

{\hspace{-0.35cm}\bf III. NEUTRINOLESS DOUBLE BETA DECAY:~} 
It is clear from the charged current interaction of this left-right gauge theory that, 
in addition to the standard contribution to $0\nu \beta \beta$ via light Majorana neutrino 
exchange, there are non-standard contributions due to the exchanges of heavy RH Majorana 
neutrinos and heavy sterile Majorana neutrinos. In addition, the extended seesaw ansatz 
manifests in non-standard contributions to lepton flavor violations and non-unitarity effects. 
In particular, we show that a dominant contribution to $0\nu \beta \beta$ arises due to mixed 
diagrams with simultaneous mediation of $W^-_L$ and $W^-_R$ bosons accompanied by light left-handed 
neutrinos and heavy right-handed Majorana neutrinos \cite{rode-ilc, spatra-jhep} as shown in Fig. 
\ref{fig:inverse-lr-mix}. We present analytic expressions for two most dominant contributions 
to the effective mass term and compare them with the standard contribution,
\begin{itemize}
\item $m^{\rm ee}_{\nu}$: which is analogous to the standard contributions, 
in this model,
\begin{eqnarray}
& & m^{\rm ee}_{\nu} = \mathcal{N}^2_{e\,i}\, m_{\nu_i} \, , 
\label{eqn:mnu-ee}
\end{eqnarray}
\item $m^{\rm ee}_{N}$: which originates from the mediation of two $W_R$'s with the 
exchange of heavy RH Majorana neutrinos,
\begin{eqnarray}
& & m^{\rm ee}_{N} =p^2\, \frac{M^4_{W_L}}{M^4_{W_R}}\,
        \frac{\left(\mathcal{R}_{e\,i}\right)^2}{M_{N_i}} \, ,
        \label{eqn:mN-ee}
\end{eqnarray}
\begin{figure}[htb]
\includegraphics[width=9cm,height=3.5cm]{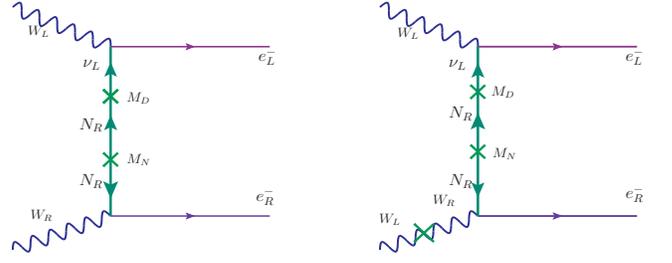}
\caption{The neutrinoless double beta decay due to $W^-_L - W^-_R$ mediation 
         in the mixed Feynman diagram with heavy and light neutrino exchanges 
         described in the text. The cross in the right panel represents left-right 
         mixing $\zeta_{LR}$.}
\label{fig:inverse-lr-mix}
\end{figure}
\item $m^{\rm ee}_{\nu N}$: which originates from simultaneous mediation of $W^-_L$ and 
$W^-_R$ and involves the Dirac mass matrix $M_D$
\begin{eqnarray}
& & m^{\rm ee}_{\nu N} \simeq  p\, \left(\zeta_{\rm LR} + \frac{M^2_{W_L}}{M^2_{W_R}}\right)\,
      \mathcal{N}_{ei} \left(M^{-1}_{N}\, M_D \, U_N\right)_{ei}\,.
      \label{eqn:mnuN-ee}
\end{eqnarray}
where, in our model, $\zeta_{\rm LR}$= LR mixing parameter $\leq 10^{-4}$.
\end{itemize}

{\hspace{-0.35cm} \bf IV. RESULTS AND DISCUSSIONS:~}
It is clear from equations (\ref{eqn:inverse-massform}) and (\ref{eqn:nustate-mass}) that 
the mass matrices $M_D$, $M$, and $M_N$ are essential for predictions of light neutrino 
masses and $0\nu \beta \beta$. At first assuming the LR gauge theory to be having its 
high scale origin from Pati-Salam symmetry and neglecting the renormalization 
group corrections, the Dirac neutrino mass matrix is approximated as the up-type quark 
mass matrix via the CKM matrix and the running masses of the three up-type quarks, 
namely, $m_u=2.33$ MeV, $m_c=1.275$ GeV, and $m_t=160$ GeV \cite{PDG} 
\begin{eqnarray}
\label{eq:inputMDckm}
&&M_D \sim V_{\rm CKM}\, \hat{M}_u\, V^T_{\rm CKM} \nonumber \\
&&\hspace{-0.8cm}= {\small \left( \begin{array}{ccc} 
0.067-0.004\,i   & 0.302-0.022\,i    & 0.55-0.53\,i   \\ 
0.302-0.022\,i   & 1.48-0.0\,i  &  6.534-0.0009\,i   \\ 
0.55-0.53\,i & 6.534-0.0009\,i    & 159.72 +0.0\, i \,.
\end{array} \right) \text{GeV}} \nonumber \\
&&
\end{eqnarray}
Under the condition $M_N \gg M \gg M_D$, the non-unitarity contribution of the extended 
seesaw model is mainly due to $\eta \simeq \frac{1}{2} X\, X^\dagger$, giving rise to 
$\eta_{\alpha \beta} = \frac{1}{2} \sum^{3}_{k=1}\, \frac{M_{D_{\alpha k}}\,M^*_{D_{
\beta k}}}{M^2_{k}}$ where we have assumed for the sake of simplicity: $M = \text{diag}
[M_1, M_2, M_3]$.  Then by saturating the available bound on $|\eta_{\tau \tau}| \leq 2.7 
\times 10^{-3} $ \cite{antu-eta, ram}, we obtain 
\begin{eqnarray}
\frac{1}{2}\, \bigg[\frac{0.5805}{M^2_1}+\frac{42.72}{M^2_2} 
                    +\frac{25510.7}{M^2_3} \bigg] = 2.7 \times 10^{-3}
\label{eqn:rel-eta}
\end{eqnarray}
where the numbers inside the square bracket are in $\text{GeV}^2$.
We note that the above relation can be satisfied in the partial degenerate case, $M_1 = 
M_2 \geq 100$ GeV, and $M_3 \geq 2.2$ TeV and also in the non-degenerate case, $M_1 \geq 
10$ GeV, $M_2 \geq 50$ GeV and $M_3 \geq 2.2$ TeV, but in the degenerate case, $M_1 = 
M_2 = M_3 = 2.2$ TeV.

{\hspace{-0.35cm}\bf a. Determination of {\bf $\mu_S$} from neutrino oscillation data:}
Inverting the neutrino mass formula given in eqn. (\ref{eqn:inverse-massform}) and   
using equation (\ref{eqn:nustate-mass}) and our model parameters, we obtain
\begin{eqnarray}
& &\mu_S\, (\,\text{GeV}\,)= X^{-1}\, \mathcal{N} \hat{m}_\nu \mathcal{N}^T\, (X^T)^{-1} \nonumber \\
&&\hspace{-0.4cm}={\small 
\left(
\begin{array}{ccc}
 0.01147+0.01 i & -0.0027-0.0024 i & 0.0007+0.002 i \\
 -0.0027-0.0024 i & 0.0006+0.0005 i & -0.0001-0.0004 i \\
 0.0007+0.002 i & -0.0001-0.0004 i & -0.00004+0.0003 i
\end{array}
\right) }\nonumber 
\end{eqnarray}
where we have used the hierarchical neutrino masses $\hat{m}_{\nu}^{\rm diag} = 
{\rm diag}(\text{0.00127 ~eV}, ~\text{0.00885 ~eV}, ~\text{0.0495 ~eV})$ and  
global fit to the neutrino oscillation data including recent values of $\theta_{13} 
= 9.0^\circ$ and $\delta=0.8\pi$ \cite{DAYA-BAY}.

Thus, in the inverse seesaw approach, the light neutrino masses and large neutrino mixings 
including non-zero values of $\theta_{13}$ can be easily fitted through the elements 
of the $\mu_S$ matrix which may have interesting consequences on leptogenesis \cite{Bdev}. 
Although we have explicitly fitted the hierarchical light neutrino masses, similar fits can 
be obtained in the inverted hierarchical as well as the quasi-degenerate cases with corresponding 
elements of $\mu_S$. In the case of $M_D$ being similar to charged lepton mass matrix which 
holds true in conventional LR gauge theories \cite{type-IImoh,moh-senj} neutrino oscillation 
data are similarly fitted with the corresponding $\mu_S$ matrix. 

{\hspace{-0.35cm}\bf b. Neutrinoless double beta decay predictions:~}
As explained in equations (\ref{eqn:nustate-mass}) -( \ref{eqn:mnuN-ee}), the mixing matrices 
$X=\frac{M_D}{M}$, $X^\prime=\frac{M_D}{M_N}$, and $X''=\frac{M}{M_N}$ all contribute to non-standard 
predictions of $0\nu \beta \beta$ amplitude in the present left-right gauge theory. We have
assumed the RH heavy Majorana neutrino mass matrix to be diagonal, $M_N = \text{diag}[M_{N_1}, M_{N_2}, M_{N_3}]$. 
Using the model parameters given in eqn. (\ref{eq:inputMDckm}) for $M_D$, $M=\text{diag}[150, 150, 2500]$ 
GeV, $M_N = \text{diag}[5000, 5000, 10000]$ GeV, $U_\nu = U_{\rm PMNS}$, $U_{N}= {\bf 1}_{3 \times 3}$, 
and $U_{S} = {\bf 1}_{3 \times 3}$, we have derived the relevant elements of the mixing matrices $\mathcal{N}_{e\, i}$, 
$\mathcal{R}_{e\, k}$, $\mathcal{U}^{\nu S}_{e\,j }$, $\mathcal{U}^{\nu N}_{e\, k}$, $\mathcal{V}^{N \nu}_{e\, i}$, 
and $\mathcal{V}^{N S}_{e\, k}$,
\begin{eqnarray}
& &\mathcal{N}_{e\, 1} = 0.819, \quad \mathcal{N}_{e\, 2} = 0.552, \quad \mathcal{N}_{e\, 3} = 0.156 \, ,\nonumber \\
& &\mathcal{R}_{e\, 1} = 0.997, \quad \mathcal{R}_{e\, 2} = 0.0, \quad \mathcal{R}_{e\, 3} = 0.0 \nonumber\, , \\
& &\mathcal{U}^{\nu S}_{e\, 1} = 0.00045, \quad \mathcal{U}^{\nu S}_{e\, 2} = 0.002, \quad \mathcal{U}^{\nu S}_{e\, 3} = 0.0002 \, , \nonumber \\
& &\mathcal{U}^{\nu N}_{e\, 1} = 0.00001, \quad \mathcal{U}^{\nu N}_{e\, 2} = 0.00005, \quad \mathcal{U}^{\nu N}_{e\, 3} = 0.000007 \, , \nonumber \\
& &|\mathcal{U}^{N \nu}| \leq 10^{-9}|\, , \quad \text{and} \quad|\mathcal{U}^{N S}| \leq 10^{-1}\, .
\end{eqnarray}

\begin{figure}[htb!]
 \includegraphics[width=8cm,height=5.0cm]{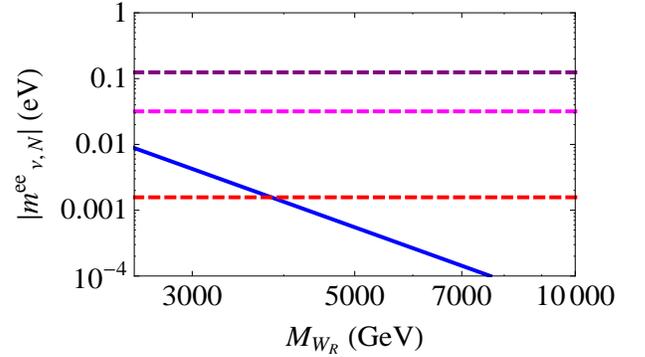}
 \caption{Contributions to effective mass $m^{\rm ee}_{\rm \nu}$ and $m^{\rm ee}_{\rm N}$ 
 as a function of $M_{W_R}$. The slanting line represents our sub-dominant contribution due 
 to $W_R-W_R$ mediation. For comparison, the standard contributions with hierarchical neutrino 
 masses ({\it the bottom horizontal line}), inverted hierarchical masses ({\it the middle horizontal line}), 
 and quasi-degenerate masses ({\it the top horizontal line}) are also given.} 
 \label{fig:subdom} 
\end{figure}

With $|p| = 100$ MeV, $M_{W_R} = 5$ TeV and using equations (\ref{eqn:nustate-mass}) -( \ref{eqn:mnuN-ee}), we predict 
the effective mass for $0\nu \beta \beta$ transition rate for hierarchical light neutrino 
masses,
\begin{eqnarray}
& &\left|m^{\rm ee}_{\nu} \right| = \mathcal{N}^2_{e\, 1}\, m_{\nu_1}+  \mathcal{N}^2_{e\, 2}\, m_{\nu_2} + 
            \mathcal{N}^2_{e\, 3}\, m_{\nu_3} \nonumber \\
& & \hspace{2.5cm}= 0.00157~ \text{eV}\, ,  
\label{eqn: LL-est}\\
& & \left|m^{\rm ee}_{N} \right|= 6 \times 10^{-4} ~ \text{eV}\, ,  
\label{eqn: RR-est}\\
& & \left|m^{\rm ee}_{\nu N} \right| \sim 1 ~\text{eV} \, .
\label{eqn: mix-est}
\end{eqnarray}
\begin{figure}[htb!]
 \includegraphics[width=8cm,height=5.0cm]{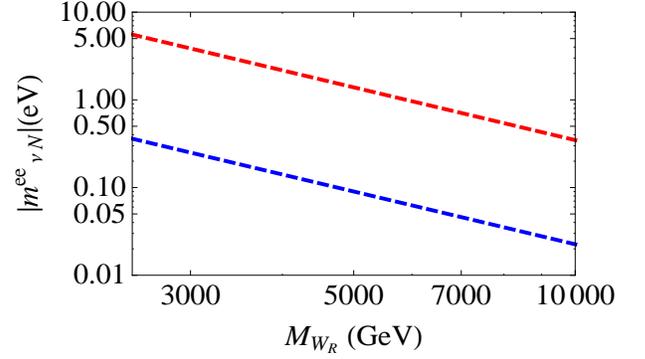}
 \caption{Contributions to effective mass $m^{\rm ee}_{\rm \nu N}$ vs $M_{W_R}$ 
 using $M_D$ similar to the up-quark mass matrix ({\it upper curve}) 
 and $M_D$ similar to charged lepton mass matrix ({\it lower curve}).}
 \label{fig:domi-mix} 
\end{figure}
Our numerical predictions are shown in Fig.\ref{fig:domi-mix} as a function of $W_R$ mass. With Dirac neutrino mass 
matrix having quark-lepton symmetric origin, the most dominant contribution due to $W^-_L$- 
$W^-_R$ mediation is found to be $m^{\rm ee}_{\nu N} \simeq$ 1 eV and $0.04$ eV for $M_{W_R} 
= 5$ TeV, and $10$ TeV, respectively. These predictions are reduced to $m^{\rm ee}_{\nu N} 
\simeq 0.07$ eV and $0.03$ eV for the corresponding values of the $M_{W_R}$ when the Dirac neutrino 
mass matrix is similar to the charged lepton mass matrix. In other words, we predict that the 
$0\nu \beta \beta$ process would be able to discriminate LR gauge models having their roots in quark-
lepton symmetry.  We note that the sub-dominant contribution due to $W^-_R$-$W^-_R$ mediation 
given in eqn. (\ref{eqn: RR-est}) is suppressed as compared to the standard contribution due to 
$W^-_L$-$W^-_L$ mediation given in eqn. (\ref{eqn: LL-est}) for the same $M_{W_R}$ masses shown 
in Fig. \ref{fig:subdom}.

For the sake of comparison with the prediction for the inverse $0\nu \beta \beta$ processes in the golden 
channel $e^-\, e^- \rightarrow W^-_L \,W^-_R$ \cite{rode-ilc} which might be phenomenologically important 
for Linear Collider searches we used $M_{W_R} \geq 2.5$ TeV \cite{WR-limit} to determine 
$\eta_\lambda = \frac{M^2_{W_L}}{M^2_{W_R}}\,\mathcal{N}_{ei} \left(M^{-1}_{N}\, M_D \, U_N\right)_{ei}$ 
which enters in cross-section for this process. Our model predicts this parameter to be $8.6 \times 
10^{-8}$ whereas the limit on this parameter is $\eta_\lambda \leq 9 \times 10^{-7}$ derived 
from the current experimental limit on $0\nu \beta \beta$ transition rate. 

{\hspace{-0.35cm}\bf c. Lepton flavor violation:~}
Besides the neutrinoless double beta decay process, the light and heavy neutrinos in this model 
can actively mediate different lepton flavor violating processes, $\mu \rightarrow e + \gamma$, 
$\tau \rightarrow e + \gamma$, and $\tau \rightarrow \mu + \gamma$ which are currently under active 
experimental investigation. The dominant contributions are mainly through the exchange of 
the six heavy neutrinos \cite{inverseothers} with branching ratio
\begin{eqnarray}
&&\text{Br}\left(\ell_\alpha \rightarrow \ell_\beta + \gamma \right) =
          \frac{\alpha^3_{\rm w}\, s^2_{\rm w}\, m^5_{\ell_\alpha}}
          {256\,\pi^2\, M^4_{W}\, \Gamma_\alpha} 
           \left|\mathcal{G}^{N}_{\alpha \beta} + \mathcal{G}^{S}_{\alpha \beta}\right|^2 
\nonumber \\ 
\label{eq:LFV}\\
&\text{where}~&\mathcal{G}^{N}_{\alpha \beta} =
        \sum_{k} \left(\mathcal{U}^{\nu\, N}\right)_{\alpha\, k}\, 
         \left(\mathcal{U}^{\nu\, N}\right)^*_{\beta\, k} 
         \mathcal{F}\left(\frac{m^2_{N_k}}{M^2_{W_L}}\right) \, ,
         \nonumber \\
& & \mathcal{G}^{S}_{\alpha \beta} = \sum_{j} \left(\mathcal{U}^{\nu\, S}\right)_{\alpha\, j}\, 
         \left(\mathcal{U}^{\nu\, S}\right)^*_{\beta\, j} 
         \mathcal{F}\left(\frac{m^2_{S_j}}{M^2_{W_L}}\right) \, , \nonumber \\
& & \mathcal{F}(x) = -\frac{2 x^3+ 5 x^2-x}{4 (1-x)^3} 
                - \frac{3 x^3 \text{ln}x}{2 (1-x)^4}\, .\nonumber 
\end{eqnarray}
Within the allowed range of model parameters  $M_N \gg M \gg M_D$, it is clear that the first 
term in eq. (\ref{eq:LFV}) is negligible. The second term involving the the heavy sterile neutrinos gives dominant 
contributions which is proportional to $\sum_{j} \left(\mathcal{U}^{\nu\, S}\right)_{\alpha\, j}\, 
         \left(\mathcal{U}^{\nu\, S}\right)^*_{\beta\, j} \simeq 2 \eta_{\alpha \beta}$ and 
our model predictions are
\begin{eqnarray}
& &\text{Br}\left(\mu \rightarrow e + \gamma \right) = 1.36 \times 10^{-15} \, ,
\nonumber \\
& &\text{Br}\left(\tau \rightarrow e + \gamma \right) = 1.06 \times 10^{-13} \, ,
\nonumber \\
& &\text{Br}\left(\tau \rightarrow \mu + \gamma \right) = 3.17 \times 10^{-12} \, .
\end{eqnarray}
Noting that the present experimental limit at 90$\%$ C.L, $\text{Br}\left(\mu \rightarrow e + 
\gamma \right) \leq 1.2 \times 10^{-11}$ \cite{LFV1} is almost three orders of magnitude 
stronger than the limits $\text{Br}\left(\tau \rightarrow e + \gamma \right) \leq 3.3 \times 10^{-8}$ or 
$\text{Br}\left(\tau \rightarrow \mu + \gamma \right) \leq 4.4 \times 10^{-8}$ 
\cite{LFV2}, appears to justify why the limit on $|\eta_{e \mu} |$ is at least one orders 
of magnitude better than the ones on $|\eta_{e \tau} |$ and $|\eta_{\mu \tau} |$. 
The projected reach of future sensitivities of ongoing searches are $\text{Br}\left(\tau \rightarrow e + 
\gamma \right) \leq 10^{-9}, ~ \text{Br}\left(\tau \rightarrow \mu + \gamma \right) 
\leq 10^{-9} $, and $\text{Br}\left(\mu \rightarrow e + \gamma \right) \leq 10^{-18}$ 
\cite{LFV3,LFV4} throughout which the model predictions can be easily verified or falsified.

{\hspace{-0.35cm}\bf V. CONCLUSION: ~}
We have shown that in a class of left-right gauge theories, the light neutrino masses naturally 
arise though gauged inverse seesaw mechanism consistent with the current neutrino oscillation data. 
The associated TeV scale masses of $W^\pm_R$ and $M_N$ can give dominant non-standard contributions to 
neutrinoless double beta decay which might be important for experimental searches. 
Specifically, we have demonstrated that the mixed diagram, via simultaneous mediation of $W^-_L$ and 
$W^-_R$ accompanied by the naturally predicted Dirac neutrino mass terms, gives the dominant 
contribution to $0\nu \beta \beta$ rate. Also this mixed diagram has rich phenomenological 
implication at ILC for the detection of the inverse process like 
$e^- e^- \rightarrow W^-_L W^-_R$. We have explicitly shown that this Dirac neutrino mass 
matrix could be similar to the up quark mass matrix which may have its high scale quark-lepton 
symmetric origin, or it may be similar to the charged lepton mass matrix expected from left-right 
gauge theory. The effective mass prediction in the former case being nearly $10$ times larger than 
the latter case, we suggest that $0\nu \beta \beta$ signatures may probe high scale quark-lepton 
symmetry. As in our approach it is not necessary to fine tune the Dirac mass matrices, the 
left-right models could serve as promising theories for charged fermion masses. The TeV scale 
masses of $W^\pm_R$ and $Z_R$ bosons are accessible to ongoing searches at LHC \cite{Kyong}. The predicted 
branching ratios for lepton flavor violating decays, being closer to the current experimental 
search limits, could be used to verify or falsify the left-right model framework considered in 
this letter. 

\begin{thebibliography}{10}
\bibitem{note} Noting that the doubly charged Higgs propagator provides additional 
damping factor to $0\nu \beta \beta$ amplitude, we ignore such contributions for 
sufficiently large values of the Higgs mass.
\end{thebibliography}
\end{document}